# Probing unexplored territories with MUSE: a second generation instrument for the VLT

Bacon R.[1], Bauer S.[2], Boehm P.[2], Boudon D.[1], Brau-Nogué S.[6], Caillier P.[1], Capoani L.[1], Carollo C.M.[3], Champavert N.[1], Contini T.[6], Daguisé E.[1], Dallé D.[1], Delabre B.[4], Devriendt J.[1], Dreizler S.[5], Dubois J.[1], Dupieux M.[6], Dupin J.P.[6], Emsellem E.[1], Ferruit P.[1], Franx M.[7], Gallou G.[6], Gerssen J.[2], Guiderdoni B.[1], Hahn T.[2], Hofmann D.[5], Jarno A.[1], Kelz A.[2], Koehler C.[5], Kollatschny W.[5], Kosmalski J.[1], Laurent F.[1], Lilly S.J.[3], Lizon J.[4], Loupias M.[1], Lynn S.[1], Manescau A.[4], McDermid R.M.[7], Monstein C.[3], Nicklas H.[5], Parès L.[6], Pasquini L.[4], Pécontal-Rousset A.[1], Pécontal E.[1], Pello R.[6], Petit C.[1], Picat J-P.[6], Popow E.[2], Quirrenbach A.[7], Reiss R.[4], Renault E.[1], Roth M.[2], Schaye J.[7], Soucail G.[6], Steinmetz M.[2], Stroebele S.[4], Stuik R.[7], Weilbacher P.[2], Wozniak H.[1], de Zeeuw P.T.[7]

(1) CRAL – Observatoire de Lyon, 9 avenue Charles André, 69230 Saint-genis-Laval, France
(2) Astrophysikalisches Institut Potsdam, An der Sternwarte 16, 14482 Potsdam, Germany
(3) ETH Zurich, Institute of Astronomy, ETH Zentrum, 8092 Zürich, Switzerland
(4) European Southern Observatory, Karl-Schwarzschild-Str. 2, 85748 Garching, Germany
(5) Institute for Astrophysics Göttingen, Friedrich-Hund-Platz 1, 37077 Göttingen, Germany
(6) LAOMP – Observatoire Midi Pyrénées, 14, avenue Edouard Belin, 31400 Toulouse, France
(7) Sterrewachte Leiden, Postbus 9513, 2300 RA Leiden, The Netherlands

Summary: The Multi Unit Spectroscopic Explorer (MUSE) is a second-generation VLT panoramic integral-field spectrograph under preliminary design study. MUSE has a field of 1x1 arcmin² sampled at 0.2x0.2 arcsec² and is assisted by the VLT ground layer adaptive optics ESO facility using four laser guide stars. The simultaneous spectral range is 0.465-0.93 µm, at a resolution of R~3000. MUSE couples the discovery potential of a large imaging device to the measuring capabilities of a high-quality spectrograph, while taking advantage of the increased spatial resolution provided by adaptive optics. This makes MUSE a unique and tremendously powerful instrument for discovering and characterizing objects that lie beyond the reach of even the deepest imaging surveys. MUSE has also a high spatial resolution mode with 7.5x7.5 arcsec² field of view sampled at 25 milli-arcsec. In this mode MUSE should be able to obtain diffraction limited data-cubes in the 0.6-0.93 µm wavelength range. Although the MUSE design has been optimized for the study of galaxy formation and evolution, it has a wide range of possible applications; e.g. monitoring of outer planets atmosphere, environment of young stellar objects, super massive black holes and active nuclei in nearby galaxies or massive spectroscopic surveys of stellar fields in the Milky Way and nearby galaxies.

## 1. Imager or spectrograph?

Imagers and spectrographs are the most common tools of optical astronomers. In most cases, astronomical observations start with imaging surveys in order to find the interesting targets and then switch to spectrographic observations in order to study the physical and/or dynamical properties of the selected object. Thanks to the excellent throughput and large format of today's detectors, large fractions of the sky can be surveyed in depth with imagers. The most limiting factor is the spectroscopic observations, which are time consuming and tend to have small multiplex capabilities. Recent development of large multi-object spectrographs such as VIMOS at VLT [LeFevre O. et al, 2003] or DEIMOS at Keck [Fabers, S. et al, 2003] has somewhat improved the situation. However the total number of sources in a typical imaging survey is much larger than what is possible to observe with spectroscopy. The selection of sources is then mandatory. Usually the selection criterion is based on a series of multi-color images and is intended to select the appropriate spectral characteristics of the population of the searched objects. This incurs a direct cost in telescope time since more than one exposure must be made at each sky location. As another disadvantage, the selection process is never 100% efficient, and thus a fraction of time of the follow-up spectroscopy is lost due to misidentifications.

The major weakness of this approach, however, is probably not the relatively low efficiency of the method, but the a priori selection of targets. This pre-selection severely biases the spectrographic observations and limits considerably the discovery space.

**2. Imager and spectrograph**

An alternative to the classical approach is to perform simultaneously imaging **and** spectroscopy. The idea is to merge into one instrument the best of the two capabilities: from the imaging world its field of view and high spatial resolution; and from the spectrograph's world its high resolving power and large spectral range.

Such an instrument will overcome the difficulty inherent to the classical method. Because there is no longer the need to pre-select the sources, one can even detect objects that would not have been found or pre-selected in the pre-imaging observations. In the most extreme case, such as object with very faint continuum but relatively bright emission lines, the objects can only be detected with this instrument, however not with direct imaging techniques.

A simple computation shows that such an ideal instrument will necessarily need a lot of detector pixels. For example let's take a spatial field of view corresponding to a standard 2048x4096 pixels detector and a wavelength range of 0.4-0.8 µm with a spectral resolution of 3000, which translate to 4000 spectral pixels. The total number of pixels is then $16 \times 10^9$. Given that some optics is needed in front of these pixels, one can immediately see the feasibility problem.

**3. MUSE wide field mode**

The Multi Unit Spectroscopic Explorer (MUSE) for the ESO/VLT telescope is a major step towards this ideal instrument. It is an integral field spectrograph (or IFU) which combines large field of view, high spatial resolution, medium resolving power and large simultaneous spectral range. Nowadays, integral field spectroscopy is part of the panoply of modern telescopes. However, most of the currently operating integral field spectrographs have only a small field of view and are thus devoted to the detailed physical study of single objects. Some multi-IFUs, like Giraffe at VLT [Pasquini et al. 2002], have multiplex capabilities of a dozen objects, which increase their efficiency. This however does not break the operational three steps (imaging, selection and spectrography) paradigm.

| | |
|---|---|
| Simultaneous spectral range | 0.465-0.93 µm |
| Resolving power | 2000@0.46µm |
| | 4000@0.93µm |
| **Wide Field Mode** | |
| Field of view | 1x1 arcmin² |
| Spatial sampling | 0.2x0.2 arcsec² |
| Spatial resolution @ 0.75µm (median seeing) | 0.46 arcsec (AO) 0.65 arcsec (non AO) |
| AO condition of operation | 70%-ile |
| Sky coverage with AO | 70% at galactic pole |
| | 99% at galactic equator |
| Limiting magnitude in 80h | $I_{AB}$ = 25.0 (full Res) |
| | $I_{AB}$ = 26.7 (R=180) |
| Limiting Flux in 80h | $3.9 \times 10^{-19}$ erg.s$^{-1}$.cm$^{-2}$ |
| **Narrow Field Mode** | |
| Field of view | 7.5x7.5 arcsec² |
| Spatial sampling | 0.025x0.025 arcsec² |
| Spatial resolution @ 0.75µm (median seeing) | 0.042 arcsec |
| Strehl ratio @ 0.75µm | 5% (10% goal) |
| Limiting magnitude in 1h | $R_{AB}$=22.3 |
| Limiting flux in 1h | $2.3 \times 10^{-18}$ erg.s$^{-1}$.cm$^{-2}$ |
| Limiting surface brightness | $R_{AB}$=17.3 arcsec$^{-2}$ |

*Table 1 : MUSE Observational Parameters*

MUSE has three operating modes: a wide field mode (WFM) with and without adaptive optics correction and a narrow field mode (NFM) with adaptive optics. The observational parameters are given in table 1.

The total number of information elements is given by the product of the number of spaxels[1] (90,000) with the number of spectral pixels (4,000), resulting in 360 million elements in the final data-cubes. Such a large number of pixels is not feasible with a single piece of optics and a single detector. MUSE is thus composed of 24 identical modules, each one consisting of an advanced slicer, a spectrograph and a $(4k)^2$ detector. A series of fore-optics and splitting and relay optics is in charge of derotating and splitting the square field of view into 24 sub-fields. These are placed on the Nasmyth platform between the VLT Nasmyth focal plane and the 24 IFU modules. AO correction will be performed by the VLT deformable secondary mirror [Arsenault R. et al, SPIE 6272-29]. Four sodium laser guide stars are used, plus a natural star for tip/tilt correction. All guide stars are taken outside the scientific field of view in order to minimize the amount of scattered light, while the only additional optic located within the scientific field of view is a Na notch filter, revolutionary reducing transmission losses with respect to traditional AO systems. This complex AO system is part of the VLT AO facility described elsewhere at this meeting [Stroebele S. et al, SPIE 6272-11, Stuik R. et al, SPIE 6272-33].
.

### 4. MUSE narrow field mode

The MUSE narrow field mode uses an additional optical system inserted into the fore-optics to change the spatial sampling from 0.2 arcsec to 0.025 arcsec. The field of view is proportionally reduced to 7.5x7.5 arcsec². The most significant change is in the AO optimization and configuration (laser guide star are moved closer) and the tip/tilt, which is performed at IR wavelengths on either a natural guide star within the field of view or the object itself. With such a configuration, the AO facility is expected to deliver a diffraction-limited image with a Strehl ratio of 5% (goal 10%) at 0.75µm.

### 5. Science case

MUSE has a broad range of astrophysical applications, ranging from the spectroscopic monitoring of solar system's outer planets to very high redshift galaxies. We give in the following sections a few examples of scientific applications that are considered to be important instrument drivers.

#### 5.1 Wide-Field Mode

The most challenging scientific and technical application, and the most important driver for the instrument design, is the study of the progenitors of normal nearby galaxies out to redshifts z>6. These systems are extremely faint and can only be found by their $Ly_\alpha$ emission. MUSE will be able to detect these in large numbers (~15,000) through a set of nested surveys of different area and depth (figure 1). The deepest survey will require very long integration (80 hrs each field, figure 2) to reach a limiting flux of $3.9 \times 10^{-19}$ erg.s$^{-1}$.cm$^{-2}$, a factor of 100 times better than what is currently achieved with narrow band imaging. These surveys will **simultaneously** address the following science goals:

- Study of intrinsically faint galaxies at high redshift, including determination of their luminosity function and clustering properties,
- Detection of $Ly_\alpha$ emission out to the epoch of reionization, study of the cosmic web, and determination of the nature of reionization,
- Study of the physics of Lyman break galaxies, including their winds and feedback to the intergalactic medium,
- Spatially resolved spectroscopy of luminous distant galaxies, including lensed objects
- Search of late-forming population III objects,
- Study of active nuclei at intermediate and high redshifts,
- Mapping of the growth of dark matter haloes,
- Identification of very faint sources detected in other bands, and
- Serendipitous discovery of new classes of objects.

---
[1] Spatial elements (to be distinguished from detector *pixels*)

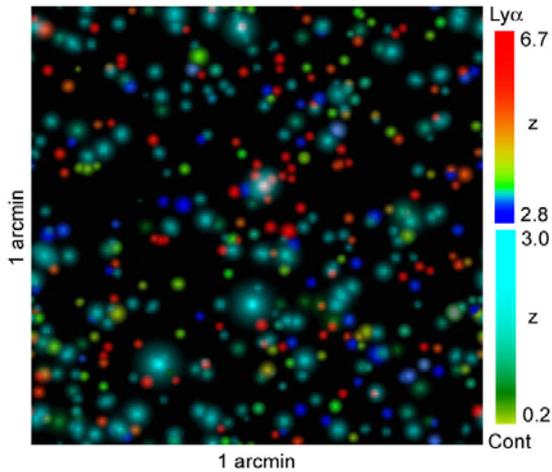

*Fig 1: Simulated MUSE deep field. Galaxies are coloured according to their apparent redshift. Galaxies detected by their continuum ($I_{AB} < 26.7$) and/or by their $Ly_\alpha$ emission (Flux > 3.9 x $10^{-19}$ erg.s$^{-1}$.cm$^{-2}$) are shown.*

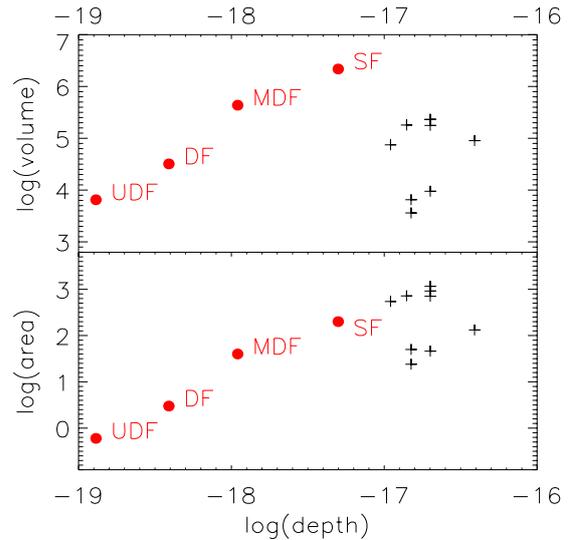

*Fig 2: Sampled area (in arcmin²) and sampled volume (comoving Mpc³) of MUSE deep fields (red circles) versus the current $Ly_\alpha$ surveys*

Multi-wavelength coverage of the same fields by MUSE, ALMA, and JWST will provide nearly all the measurements needed to answer the key questions of galaxy formation.

At lower redshifts, MUSE will provide exquisite two-dimensional maps of the kinematics and stellar populations of normal, starburst, interacting and active galaxies in all environments, probing sub-kiloparsec scales out to well beyond the Coma cluster. These will reveal the internal substructure, uncovering the fossil record of their formation, and probe the relationship between super massive black holes and their host galaxy.

MUSE will enable massive spectroscopy of the resolved stellar populations in the nearest galaxies, outperforming current capabilities by factors of over 100. This will revolutionize our understanding of stellar populations, provide a key complement to GAIA studies of the Galaxy, and a preview of what will be possible with an ELT (Fig.3).

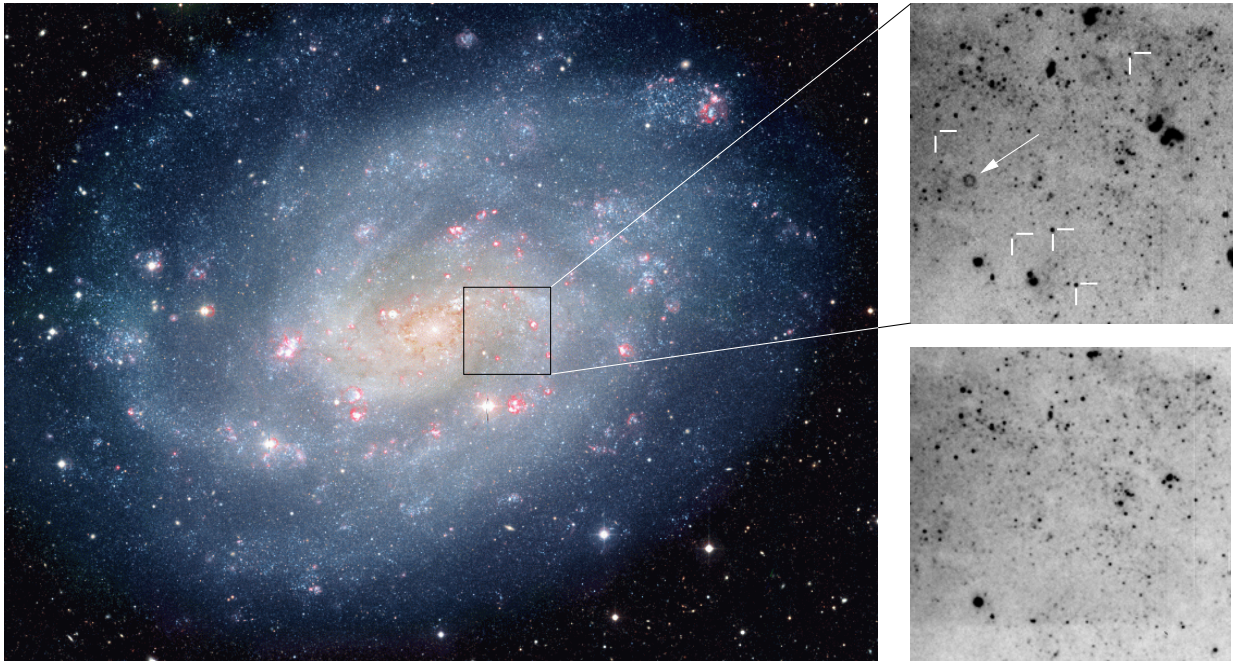

*Fig. 3: Left: composite image of the southern spiral galaxy NGC300, illustrating the power of massive spectroscopy with MUSE. The frames to the right are a narrowband [O III] 5007 exposure (top), and a corresponding nearby continuum exposure (bottom), obtained with the NTT over a FOV of 2.2×2.2arcmin$^2$ (Soffner et al. 1996). MUSE will cover the same field in a total of 4 exposures. Unlike the narrowband imaging example (intended for the purpose of discovering planetary nebulae), the MUSE datacube will provide full spectral information for each spatial element, with a huge discovery potential for massive stars, super bubbles, H II regions, PNe, SNRs, novae — virtually the full inventory of the stellar and gaseous constituents of the galaxy. As a complement to GAIA, application to LMC/SMC and the bulge of the MW will provide kinematics and abundance information for the detailed study of stellar populations and the formation history of the host galaxy, similar to e.g. the RAVE survey (Steinmetz 2002), albeit orders of magnitude more efficiently.*

### 5.2 Narrow-Field Mode

Contrary to the Wide Field Mode, the Narrow Field mode science is dedicated to detailed study of single objects at very high spatial resolution. We give in the following a few examples.

The study of super massive black holes: During galaxy mergers, super massive black holes sink to the bottom of the potential well, forming binary systems which 'scour out' lower-density cores in the central regions of the remnant. Such processes should leave detectable signatures in the environment of the SMBH. Likewise, accretion of mass onto super massive black holes should trigger activity and feedback to the local regions and beyond. However, observationally, very little is known about this environment, either in terms of stellar orbital structure or chemical enrichment history.

Young stellar objects: The key contribution from MUSE will be both in spectral grasp (covering key diagnostics of density, temperature and ionization) and the ability to provide very high spatial resolution over a relatively large field of view. This will allow the physical processes involved in the formation and structure of the jets to be investigated in detail

Solar system: MUSE NFM would allow observation of various bodies within our solar system at a spatial resolution approaching that of more costly space missions. Applications are: monitoring volcanic activity on the Galilean satellites, spectral monitoring of Titan's atmosphere, global monitoring of the atmospheres of Uranus and Neptune, internal structure and composition of comets and mineralogical surface heterogeneities of asteroids.

## 6. Opto-mechanical concept

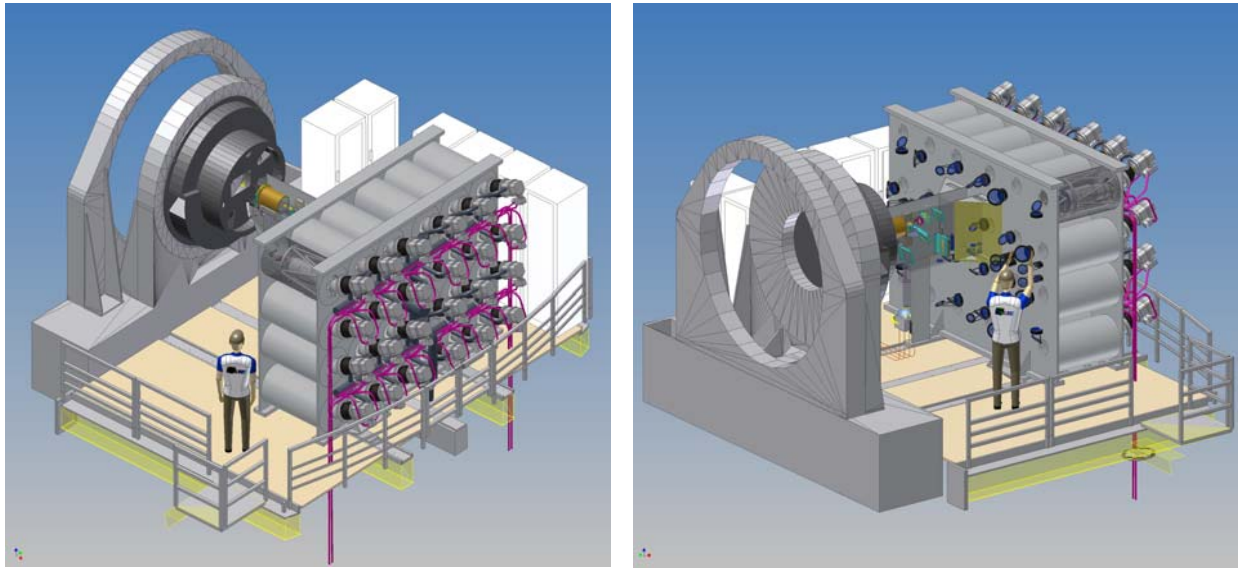

*Fig 4: Instrument overview at the VLT Nasmyth platform.*

The opto-mechanical concept has to fulfill the following challenging requirements:
- Replication of modules at low cost in order to achieve the required number of spatial and spectral elements.
- High throughput despite the required number of optical surfaces
- High image quality in order to optimally use the image quality deliver by the AO facility
- High stability and reliability over long exposures
- Maintain cost, mass and volume

The 24 IFUs are central to MUSE. They have been designed to achieve an excellent image quality (85% enclosed energy within a 15x30µm$^2$ in the detector plane), and make use of innovative slicer and spectrograph concepts. The slicer is an evolution of the advanced slicer concept proposed by R. Content [Content R. et al, 1997]. It is based on a 2-mirror compact design, suitable for diamond machining. Recent progress of the manufacturing process has enabled high precision metal surfacing with good surface roughness (3 nm RMS). Such mirrors are now compatible with optical wavelength requirements and are much more cost effective than other approaches for the large-scale production foreseen for MUSE. Further details of the MUSE slicer are presented elsewhere at this conference [Laurent F. et al, SPIE 6273]. The compact spectrograph design achieves an excellent image quality over the large spectral bandwidth of MUSE. In this design, the tilt of the detector compensates for the axial chromatism, which then does not need to be corrected optically. This is a cost effective solution, avoiding the use of expensive optical materials, e.g. CaF2.

To maintain a high throughput (40% for the whole instrument) despite the relatively large number of required surfaces, attention is paid to use state-of-the art transmission and reflection coatings. Detectors are 4kx4k 15µm deep depletion devices with improved quantum efficiency in the red. Furthermore we will use new volume phase holographic gratings with a high efficiency over the large (one octave) spectral range.

To simplify the interfaces between GALACSI and MUSE, all AO components, including the tip/tilt sensor, are mounted in the Nasmyth derotator. There is however a risk of misalignment of the AO reference system with respect to MUSE, which is located on the platform. To mitigate this risk and to maintain the optical axis within the tight tolerances required by the spatial performances and stability, a metrology system has been designed. It is a close loop system based on four reference light sources located in the fore-optics and imaged into the AO system.

The cryogenic system is based on pulse tubes, which are compact and which avoid to refill 24 dewars with liquid nitrogen. The accompanying compressors are located outside the Nasmyth platform on the telescope floor which avoids any possible transmittance of vibrations onto the instrument.

The instrument weight is approaching 8 metric tonnes in total and its size will fill basically the entire volume of the Nasmyth platform of roughly 50 $m^3$. This outnumbers every instrument that was built so far for the VLT and will make of MUSE an impressive instrument (Fig.4). With these dimensions, assembling and providing the necessary access to all the components is a challenge. The main instrument structure is designed as a single unit to fulfil the highly demanded stability of all optical components regarding each other in order to maintain the superb image quality given by GALACSI on long exposures.

The latter is done with a complex optical system that has to derotate and to split the observing field and to distribute and feed the spectrographic units with these sub-fields. Despite its 24 spectrographs mounted into a monolithic structure, MUSE will act as a single instrument with respect to the telescope and the AO system. Nevertheless, the instrument is set up with a highly modular character for the assembly, maintenance and any operational exchange.

**7. Operation and data reduction**

Despite its impressive number of optomechanical elements, MUSE shall be an instrument easy to operate. There are no moving parts in the 24 modules and the switch between wide to narrow field mode implies only the addition of some optics within the fore-optics train. MUSE has only three operating modes: non-AO and AO wide field mode, and AO narrow-field mode. The three modes differ only by the presence of AO and the spatial sampling. In the wide field non-AO mode, operation shall be limited to the simple point-and-shoot scheme. In the other modes, the complexity is related to the operation of AO including the lasers. All modes share the same spectroscopic configurations (wavelength range and resolution).

On the other hand, with 1.6 Gb per single exposure, the data reduction is a challenge, not only because of this data volume, but also because of its 3D characteristics. The handling of such large data cubes is not straightforward. As an example, one can mention the optimal summation of a series of data cubes obtained with AO and different atmospheric conditions. This is intrinsically a 4-dimensional problem because the AO-delivered PSF changes with time, location within the field of view, and wavelength.

**8. Project status**

The MUSE Consortium consists of groups at Lyon (PI institute, CRAL), Göttingen (IAG), Potsdam (AIP), Leiden (NOVA), Toulouse (LAOMP), Zurich (ETH) and ESO. The project is currently in its preliminary design phase. In July 2006, the optical preliminary design review will be the starting point for the manufacturing of a complete breadboard consisting of a slicer, a spectrograph and a detector, while the full preliminary design review is scheduled for early 2007. Results of the breadboard will be analyzed for the final design review in July 2008. Manufacturing, assembly and integration will then take place up to mid 2011. First light is scheduled on Paranal early 2012.

**9. Conclusions**

Astronomy is to a significant degree still driven by unexpected discovery (e.g. dark matter and dark energy). These discoveries are often made by pushing the limit of observations with the most powerful telescopes and/or opening a new area of instrumental parameter space. MUSE is designed to push the VLT to its limit and to open a new parameter space area in sensitivity, spatial resolution, field of view and simultaneous spectral coverage. We are convinced that it fulfils all the required conditions to have a large potential of discoveries:

- It will be the first spectrograph that could blindly observe a large volume of space, without any imaging pre-selection.
- It will be the first optical AO-assisted IFU working at improved spatial resolution in most atmospheric conditions with large sky coverage.
- It will be the first spectrograph optimized to work with very long integration times and to reach extremely faint emission line detection.

MUSE will thus be able to discover objects that have measurable emission lines, but with a continuum that is too faint to be detected in broad-band imaging. For example, the deepest broad-band imaging available today is the HST Ultra Deep Field (UDF) with $I_{AB}<29$. According to CDM simulations, however, only 15% of MUSE high-z $Ly_\alpha$ emitters (z>5.5) will have a continuum bright enough to be detected in the UDF. MUSE is also the only instrument capable of detecting faint diffuse ionized gas, like extended halos or filaments. Finally, objects with unusual spectral features should also be detected by MUSE, whatever their broad band magnitude and colors are. The unprecedented capabilities of MUSE should also lead to discoveries far away from our present expectations.

By many aspects, MUSE is a precursor of future ELT instrumentations. For example manufacturing, integration and maintenance of a large number of identical, high performance optical systems at low cost and on reasonable time scale will be a critical aspect fro most of ELT instruments. MUSE will also be the first AO-assisted IFU to address a key science case of future ELTs: massive spectroscopy of resolved stellar populations in nearby galaxies, employing crowded field 3D spectroscopy over a large field-of-view.

*MUSE public web site:* http://muse.univ-lyon1.fr